\documentclass[lettersize,journal]{IEEEtran}
\usepackage{amsmath,amsfonts}
\usepackage{amssymb, amsthm}
\usepackage{array}
\usepackage{graphicx}
\usepackage{cite}
\usepackage{enumitem}
\usepackage{float}
\usepackage{upgreek}
\usepackage{bm}
\usepackage{stfloats}
\usepackage{placeins}
\usepackage[nospread,checkfloat]{flushend}
\usepackage{hyperref}
\usepackage{threeparttable}
\usepackage{xcolor}
\theoremstyle{plain}
\newtheorem{proposition}{Proposition}

\begin{document}

\title{IMU-Aided Motion Compensation of RSS Measurements for Dynamic Visible Light–Inertial Localization}

\author{Xiao Sun, Zhenqi Zheng, Jianzhu Huai, Xingxing Li, Wanwei Zhang, Yuan Zhuang
\thanks{Corresponding author is Zhenqi Zheng (e-mail: zhengzhenqi@whu.edu.cn).}
\thanks{Xiao Sun, Xingxing Li, and Wanwei Zhang are with School of Geodesy and Geomatics, Wuhan University, Wuhan, China. Xiao Sun is also with State Key Laboratory of Precision Geodesy, Innovation Academy for Precision Measurement Science and Technology, CAS, Wuhan, China.}
\thanks{Zhenqi Zheng, Jianzhu Huai, and Yuan Zhuang are with State Key Laboratory of Information Engineering in Surveying, Mapping and Remote Sensing, Wuhan University, Wuhan, China.}
}

\markboth{}%
{Sun \MakeLowercase{\textit{et al.}}: IMU-Aided Motion Compensation of RSS Measurements}


\maketitle

\begin{abstract}
\boldmath
Visible Light Positioning (VLP) offers high-precision indoor positioning using existing LED infrastructure. However, in VLP systems using Photodiodes (PDs) as receivers, rapid movement of the PD causes measurement inaccuracies as the Discrete Fourier Transform (DFT) processing yields averaged rather than instantaneous Received Signal Strength (RSS) values. To address this challenge, we propose an RSS correction algorithm that integrates Inertial Measurement Unit (IMU) data with VLP observations to compensate for motion-induced distortions. Our approach leverages IMU-derived motion parameters to model the temporal variation of optical signals during the DFT window, enabling instantaneous RSS estimation. The algorithm is embedded within both tightly and simply coupled VLP/INS integration frameworks that estimate the pose of the moving receiver. This work targets highly dynamic indoor navigation. Simulations validate the correction model, and total-station-tracked experiments confirm reduced RSS discrepancies against model references. Across eight held-out high-dynamic handheld trials with peak translational and angular speeds of $2.25~\mathrm{m/s}$ and $284^\circ/\mathrm{s}$, correction reduces the pooled 3D position and PD-normal inclination root mean square errors relative to the LiDAR/IMU reference with a shared IMU by 56.0\% and 71.9\%, respectively. Both datasets and source code are publicly available.
\end{abstract}


\section{Introduction}
\label{introduction}

Visible Light Positioning (VLP) has emerged as a promising technology for high-accuracy indoor localization, leveraging existing Light-Emitting Diode (LED) lighting infrastructure to simultaneously provide illumination and positioning signals. Compared with widely used Indoor Positioning System (IPS) methods nowadays such as WiFi, Bluetooth, Ultra-WideBand (UWB), odometer, vision, and LiDAR, VLP shows multiple advantages \cite{8292854,el2021indoor,davidson2016survey} at the same time: high precision, high bandwidth, long lifetime, and low infrastructure cost \cite{sun2025sana}.

Among existing literature on LED-based IPS, photodiodes (PDs) and cameras serve as the primary receiving devices in VLP systems. This study focuses on PD-based VLP systems. Within PD-based frameworks, Received Signal Strength (RSS) represents the most straightforward measurement readily acquired from a single PD. Multiple PD arrays enable Angle of Arrival (AOA) measurements \cite{8675983, 9456850}, while supplementary equipment facilitates Time of Arrival (TOA) \cite{7339418} and Time Difference of Arrival (TDOA) \cite{8368233} positioning. In contrast, image-based VLP systems, despite the widespread integration of cameras in modern devices \cite{9330552,10379113}, rely on imaging principles and typically entail higher hardware and computational costs.

Although VLP achieves high positioning accuracy with low cost, it suffers from sensitivity to light blockages and receiver inclinations. Integrating VLP with an Inertial Navigation System (INS) offers a solution to these challenges \cite{ZHUANG202362}. Recent studies have explored both tightly \cite{9184060, kim_single_2022, sun2025sana, 10938845} and loosely \cite{7997704, 8322671,li_crowdfusion_2023} coupled VLP/INS systems. While both approaches leverage IMU benefits, tightly coupled models better resist IMU errors and provide continuous navigation during blockages. Various implementations exist: 2D systems assuming level placement \cite{9184060}, PD position fusion with INS \cite{8322671}, 2D PDR fusion \cite{7997704, kim_single_2022,li_crowdfusion_2023}, and smartphone ambient light sensor fusion \cite{zhuang_visible_2024}. Sun et al. \cite{sun2025sana} proposed a 6-DoF tightly coupled VLP/INS system using graph optimization, while Wang et al. \cite{10938845} developed a data-model dual-drive system. VLP can also be integrated with INS and GNSS for seamless indoor/outdoor positioning \cite{SUN2025102781}.

In highly dynamic scenarios such as UAV flight or rapid hand-held device movement, sensor data acquisition often suffers from various forms of distortion. In camera-based systems, rapid motion can cause image blurring. GIVL-SLAM compensates LiDAR scan distortion using IMU-derived poses before tightly coupled factor-graph estimation~\cite{10552105}. In the context of VLP based on highly dynamic platforms, most existing studies have focused on image sensor systems, primarily addressing issues such as image blurring and high latency caused by rapid translation and rotation. Guan et al. \cite{8368189} proposed a novel VLP algorithm based on optical flow detection and Bayesian prediction to mitigate image blurring induced by high-speed movement. Lin et al. \cite{Lin20} employed lightweight image processing algorithms to reduce latency in VLP systems during fast motion. Li et al. \cite{photonics9090632} introduced a deep learning framework based on convolutional neural networks to improve the success rate of decoding user identification information from VLP-LED signals. However, these studies did not address whether their proposed algorithms are applicable to PD-based systems, nor did they suggest how such algorithms could be adapted for PD systems. Here we focus on the motion-induced bias of windowed RSS measurements from a single moving PD.

\begin{figure}[H]
	\centering
	\includegraphics[width=0.8\linewidth]{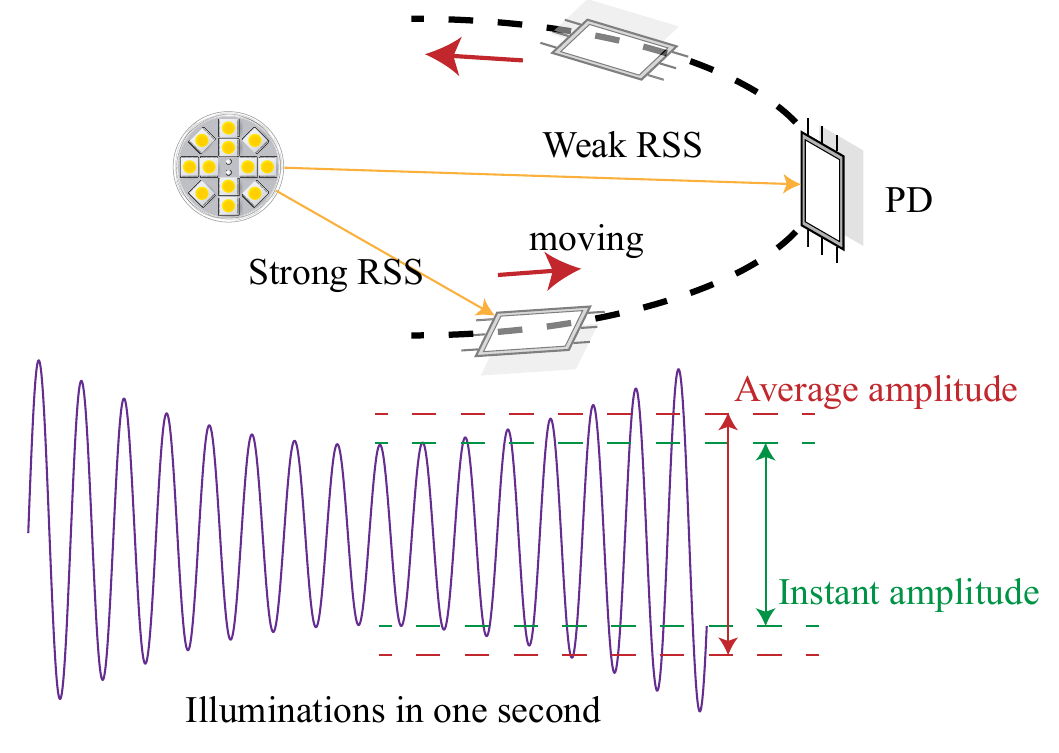}
	\caption{Error sources during rapid translation and rotation of the object.}
	\label{principle}
\end{figure}

As illustrated in Fig.~\ref{principle}, consider a scenario where an LED emits a sinusoidally modulated optical signal, and a PD rapidly traverses an arc-shaped trajectory within one second. The received optical signal at the PD thus manifests as a sinusoidal waveform with time-varying amplitude, reflecting the changing distance and incidence angle between the LED and the moving receiver. Notably, the RSS exhibits an inverse relationship with distance: weaker signals are received when the PD is farther from the LED, while stronger signals are detected when the PD approaches the LED.

In a VLP system, RSS measurements are typically obtained by applying a Discrete Fourier Transform (DFT) to raw photocurrent data collected over a time window \cite{10839122}. However, during the rapid translation and rotation of a vehicle-mounted PD, changes in position and orientation within this interval cause fluctuations in the instantaneous optical intensity. When this non-stationary signal is processed via a DFT window (typically a one-second interval), the resulting spectral component corresponds to an averaged amplitude rather than the instantaneous optical intensity (shown in Fig.~\ref{principle}). Consequently, the systematic discrepancy between the estimated and true instantaneous RSS degrades positioning accuracy. This study refines the conventional Lambertian model to account for temporal RSS variation within the DFT window and uses inertial data to correct motion-induced RSS distortions before fusion. The contributions are as follows:

\begin{itemize}
	\item{\textbf{Dynamic RSS Formation Model}: We identify and formulate a previously unmodeled error source in PD-based VLP: DFT demodulation returns a window-averaged RSS, whereas mobile positioning estimators consume it as an instantaneous RSS. This modeling assumption is different from the static per-epoch RSS assumption used in \cite{sun2025sana}.}
	\item{\textbf{IMU-aided Incremental Correction}: We derive an RSS correction term that maps high-rate IMU-derived translation and rotation inside the Fourier window to the difference between averaged and midpoint RSS. The module is estimator-agnostic and can be inserted before either the tightly coupled graph of \cite{sun2025sana} or a simple VLP/INS integration pipeline.}
	\item{\textbf{Open High-Dynamic Datasets and VLP/INS Fusion Code}: To the best of our knowledge, we are the first to publicly release high-dynamic VLP datasets containing synchronized PD and IMU measurements together with VLP/INS fusion code, including the proposed RSS correction. The datasets cover rapid attitude changes and fast running, enabling reproducible evaluation of PD-based VLP/INS under highly dynamic motion.}
\end{itemize}

Compared with our earlier work \cite{sun2025sana}, which focuses on tightly coupled pose estimation using epoch-wise Lambertian RSS residuals, this work models and corrects the motion-induced difference between window-averaged and midpoint RSS before fusion. The correction uses high-rate IMU measurements without adding states or graph factors and can also be applied to simple VLP/INS integration.

\section{Visible Light Positioning aided with INS}
This section outlines the system, optical model, and two VLP/INS integration schemes used with RSS correction.

\subsection{System Overview}
\label{overview}

\begin{figure}
	\centering
	\includegraphics[scale=0.5]{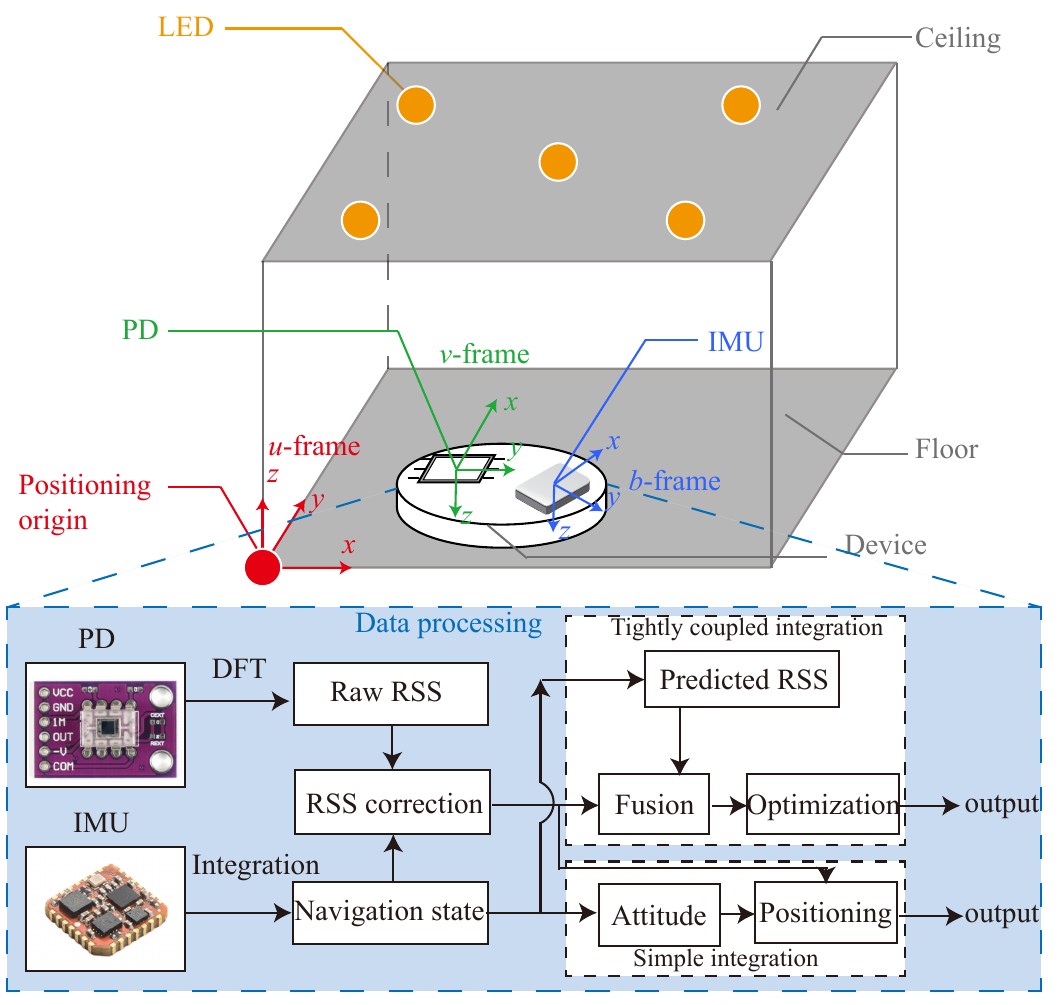}
	\caption{System architecture and data processing pipeline.}
	\label{coordinate}
\end{figure}

Fig.~\ref{coordinate} shows the LEDs, PD, IMU, and processing pipeline. DFT separates the LED signals to obtain RSS measurements, while IMU integration provides motion estimates for RSS correction. The corrected measurements enter either the tightly coupled or simple VLP/INS integration scheme to estimate the receiver pose.

Superscripts $u$, $b$, and $v$ denote the indoor, IMU body, and VLP frames, respectively (red, blue, and green in Fig.~\ref{coordinate}). The indoor frame has an arbitrarily chosen origin and orthogonal axes; the body frame follows the IMU. The VLP frame is defined by the PD plane and forward direction, with forward-right-down axes. The symbols $b_k$ and $v_k$ denote the corresponding frames at $t_k$.

\subsection{Quaternion-based Lambertian Model}
\label{lambertian}
\begin{figure}
	\centering
	\includegraphics[width=0.3\textwidth]{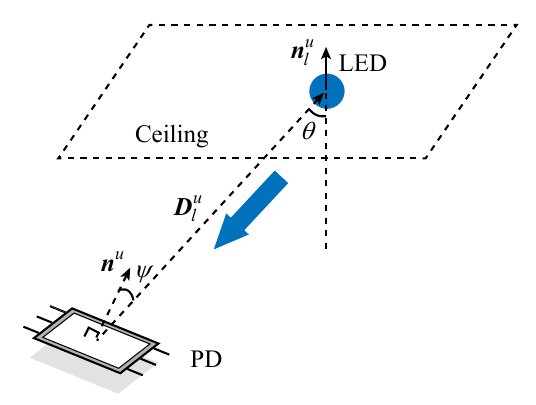}
	\caption{The signal transmitting process in VLP \cite{sun2025sana}.}
	\label{angle}
\end{figure}

For $N$ LEDs and a single PD, DFT demodulation separates the known LED modulation frequencies and extracts the RSS $P_l$ of LED $l$. With irradiance angle $\theta$ and incidence angle $\psi$ defined in Fig.~\ref{angle}, the received optical power follows the Lambertian model~\cite{sun2025sana}:
\begin{equation}
	\label{lamber2}
	P_l=\frac{\left(m_l+1\right) A_\mathrm{R} P_{\mathrm{T}l}}{2 \mathrm{\uppi}} \cdot \frac{ \left(\bm{n}^u\cdot\bm{D}_l^u\right) \left(\bm{n}_{l}^u\cdot\bm{D}_l^u\right)^{m_l}} {D^{3+m_l}_l}
\end{equation}
where $A_\mathrm{R}$ is the effective PD area, $P_{\mathrm{T}l}$ and $m_l$ are the transmit power and Lambertian order of LED $l$, and $\bm{D}_l^u$ is the PD-to-LED vector with magnitude $D_l$. The PD normal $\bm{n}^u$ and the unit vector $\bm{n}_l^u$ opposite to the LED emission direction are both defined upward. For ceiling-mounted LEDs and an upward-positive indoor $z$-axis, $\bm{n}_l^u=[0,0,1]^{\mathrm T}$; it is transformed for other frame orientations.

Receiver attitude, represented by a rotation matrix $\bm{C}$ or Hamilton quaternion $\bm{q}$, determines the PD normal:
\begin{equation}
	\label{npd}
	\bm{n}^u=\bm{C}^u_{v}\bm{n}^{v}
\end{equation}
where $\bm{C}^u_v$ is the direction cosine matrix (DCM) from $v$-frame to $u$-frame and $\bm{n}^v=[0,0,-1]^{\mathrm T}$. The Hamilton quaternion $\bm{q}^u_v=q_0+q_1\mathbf{i}+q_2\mathbf{j}+q_3\mathbf{k}$ gives the same rotation through the standard quaternion-to-DCM conversion.

\subsection{Simple Integration of VLP and INS}
\label{simple integration}
In the single-PD Lambertian model (\ref{lamber2}), receiver attitude enters RSS only through the unit normal $\bm{n}^u$, which has two independent orientation parameters. At a fixed PD position, rotation about this normal leaves all instantaneous RSS observations unchanged, so increasing the LED count cannot recover all six pose degrees of freedom from instantaneous RSS alone. As shown in Fig.~\ref{coordinate}, simple VLP/INS integration substitutes the INS-derived PD normal into the Lambertian model and solves for the three position unknowns using the available LED observations, subject to adequate LED geometry.

This approach relies on the accuracy of the INS-derived PD normal; errors in that normal propagate through the Lambertian model into the position solution.

\subsection{Tightly Coupled VLP/INS Positioning}
\label{integration}
We use the tightly coupled VLP/INS framework of \cite{sun2025sana} as a backend and baseline. It jointly estimates position and attitude from epoch-wise RSS and IMU pre-integration~\cite{RN297} over successive intervals $[t_k,t_{k+1}]$. The original backend treats DFT-derived RSS as instantaneous epoch measurements. Our module corrects RSS before residual construction, leaving the state vector and graph structure unchanged.

This VLP/INS integration scheme uses a 16-dimensional vector to form the state,
\begin{equation}
	\bm{x}=\left(\begin{matrix}(\bm{p}^u)^{\mathrm T} & (\bm{v}^u)^{\mathrm T} & (\bm{q}^u_{v})^{\mathrm T} & (\bm{b}_{a}^{v})^{\mathrm T} & (\bm{b}_{g}^{v})^{\mathrm T}\end{matrix}\right)^{\mathrm T}
\end{equation}
where $\bm{p}^u$, $\bm{v}^u$, $\bm{q}^u_v$, $\bm{b}_a^v$, and $\bm{b}_g^v$ denote position, velocity, attitude quaternion, accelerometer bias, and gyroscope bias. Before pre-integration, the fixed installation DCM $\bm{C}_b^v$ transforms the accelerometer specific-force measurements $\hat{\bm{a}}_i^b$, gyroscope angular-rate measurements $\hat{\boldsymbol{\omega}}_i^b$, and their biases into $v$-frame. The pre-integrated measurement $\hat{\bm{z}}_{v_{k+1}}^{v_k}=\left[\hat{\boldsymbol{\alpha}}_{v_{k+1}}^{v_k},\hat{\boldsymbol{\beta}}_{v_{k+1}}^{v_k},\hat{\boldsymbol{\gamma}}_{v_{k+1}}^{v_k}\right]$ comprises the pre-integrations of position, velocity, and attitude, respectively~\cite{RN297}.

For the sliding-window states $\boldsymbol{\mathcal{X}}=[\bm{x}_0,\bm{x}_1,\ldots,\bm{x}_n]$, let $\bm{r}_{\mathcal{B}}(\hat{\bm{z}}_{v_{k+1}}^{v_k},\boldsymbol{\mathcal{X}})$ denote the IMU pre-integration residual and $\boldsymbol{\varSigma}_{v_{k+1}}^{v_k}$ its covariance. Their definitions follow \cite{sun2025sana,RN297}.

Following \cite{sun2025sana}, the VLP residual $\bm{r}_{\mathcal{V}}\left(\hat{P}_l^k,\boldsymbol{\mathcal{X}}\right)$ is the Lambertian RSS predicted by (\ref{lamber2}) from $\boldsymbol{\mathcal{X}}$ minus the RSS measurement $\hat{P}_l^k$ of LED $l$ at epoch $k$. The sliding window advances as new measurements arrive.

The geometric quantities $\bm{D}_l^u$, $\bm{n}^u$, and $D_l$ follow the state-dependent model in Section~\ref{lambertian}. The PD position includes the IMU-to-PD lever arm:
\begin{equation}
	\label{sec4:leverarm}
	\hat{\bm{p}}^u = \bm{p}^u + \bm{C}^{u}_v\bm{C}^{v}_b \boldsymbol\ell^{b}
\end{equation}
where $\boldsymbol\ell^{b}$ is the lever-arm from the IMU center to the PD center in the $b$-frame.

At each window, nonlinear least squares minimizes the squared Mahalanobis norms of IMU, RSS, and constraint residuals:
\begin{equation}
	\label{cost}
	\begin{aligned}
		\min \limits_{\boldsymbol{\mathcal{X}}}&\left\{\sum\limits_{k=0}^{n-1}\left\|\bm{r}_{\mathcal{B}}\left(\hat{\bm{z}}_{v_{k+1}}^{v_k}, \boldsymbol{\mathcal{X}}\right)\right\|_{\boldsymbol{\varSigma}_{v_{k+1}}^{v_k}}^2\right.\\
		&+\sum\limits_{k=0}^{n} \sum\limits_{l=1}^{N}\left\|\bm{r}_{\mathcal{V}}\left(\hat{P}_l^k, \boldsymbol{\mathcal{X}}\right)\right\|_{\varSigma_l}^2\\
		&+\left.\sum\limits_{k=0}^{n} \left\|\bm{r}_{\mathcal{C}}\left(\bm{z}_{k}^u, \boldsymbol{\mathcal{X}}\right)\right\|_{\varSigma_k}^2\right\}
	\end{aligned}
\end{equation}
where $\varSigma_l$ is the RSS covariance of LED $l$ and $N$ is the number of available LEDs per epoch. Here, $\bm{z}_k^u$ denotes the constraint observation at epoch $k$, and $\varSigma_k$ its covariance. The constraint residual $\bm{r}_{\mathcal{C}}(\cdot)$ may represent non-holonomic constraints (NHC) for robots without lateral velocity~\cite{7247673} or height constraints for 2D positioning~\cite{sun2025sana}.

Backend details follow \cite{sun2025sana}; the backend and RSS correction implementation are available at \url{https://github.com/ShawnSun95/OB_GINS_VLP}. The following section develops the RSS correction.

\section{RSS Measurement Correction in VLP}
\label{VLP}
To address the issue that DFT processing yields time-averaged rather than instantaneous RSS values in highly dynamic VLP systems, this section presents a novel RSS correction algorithm that leverages IMU data to compensate for motion-induced distortions. The core idea is to utilize high-frequency IMU measurements to estimate the difference between the time-averaged RSS from DFT and the true instantaneous RSS values, enabling correction of the temporal averaging effect that degrades positioning accuracy in dynamic scenarios.

\subsection{Differential Form of Lambertian Model}

To implement RSS correction, it is necessary to first investigate the differential properties of the Lambertian model. By taking full derivatives of equation (\ref{lamber2}) with position disturbance $\mathrm{d}\bm{r}$ and angle disturbance $\mathrm{d}\boldsymbol{\phi}$, it has \cite{sun2025sana}:
\begin{equation}
	\label{derivative}
	\begin{aligned}
		\mathrm{d}P_l=&-P_l\frac{\bm{D}_l^u\times\bm{n}^u} {\bm{D}_l^u\cdot\bm{n}^u} \cdot \mathrm{d}\boldsymbol{\phi} + P_l\left[-\frac{\bm{n}^u}{\bm{n}^u\cdot\bm{D}_l^u} \right.\\
		&-\frac{m_l\bm{n}_{l}^u}{\bm{n}_{l}^u\cdot\bm{D}_l^u}+\left.(3+m_l)\frac{\bm{D}_l^u}{D^{2}_l}\right] \cdot \mathrm{d}\bm{r}
	\end{aligned}
\end{equation}

Equation~(\ref{derivative}) reuses the Lambertian differential form from \cite{sun2025sana} with a sign convention adjustment: we add a negative sign to the definition of $\mathrm{d}\boldsymbol{\phi}$ because \cite{sun2025sana} defines it as $\mathrm{d}\boldsymbol{\phi}^u_{vu}$, whereas $\mathrm{d}\boldsymbol{\phi}^u_{uv}$ is more commonly used in the literature to represent attitude perturbations. This differential model is only the starting point. The new derivation below uses it to integrate RSS variation across the entire DFT window and to recover the midpoint RSS required by a positioning estimator.

\subsection{Mathematical Foundation of DFT Averaging Effect}
\label{lemma1}

As illustrated in Fig.~\ref{principle} and discussed in Section~\ref{introduction}, due to the motion of the PD, the RSS value obtained via the DFT generally fails to represent the instantaneous RSS and thus requires correction. To this end, we first present the following proposition:

\begin{proposition}[DFT Averaging]
	For a nonnegative amplitude envelope and a modulation tone aligned with a DFT bin, the normalized DFT amplitude approximates the window-averaged envelope when the envelope projection at twice the modulation frequency is negligible.
	\label{prop1}
\end{proposition}
For a bin-aligned signal $x(n)=A(n)\sin(2\pi k_0n/N)$ with $A(n)\geq0$, let $X(k_0)$ denote its DFT coefficient over a rectangular window of $N$ samples. The normalized amplitude satisfies
\begin{equation}
    A_{\mathrm{DFT}}=\frac{2|X(k_0)|}{N}\approx\bar{A}=\frac{1}{N}\sum_{n=0}^{N-1}A(n).
\end{equation}
The approximation error is bounded by the envelope projection at twice the modulation frequency:
\begin{equation}
    \begin{aligned}
        B_{\mathrm{rect}}&=\left|\frac{1}{N}\sum_{n=0}^{N-1}A(n)e^{-j4\pi k_0n/N}\right|,\\
        |A_{\mathrm{DFT}}-\bar{A}|&\leq B_{\mathrm{rect}}.
    \end{aligned}
\end{equation}
Thus, the averaging approximation requires bin alignment and a small residual envelope projection. With tapered windows, the corresponding average is window-weighted. The complete proof is provided in Section~I-A of the supplementary material.

Controlled single-tone tests show close agreement between DFT-derived amplitudes and the corresponding window averages, with errors bounded by the neglected projection term. These results support the averaging approximation within the tested settings; the complete parameter sweeps and projection-bound verification are provided in Section~I-B of the supplementary material.

\begingroup
\raggedbottom
\allowdisplaybreaks[4]
\subsection{RSS Correction}

Conventionally, the RSS obtained from the DFT window is incorrectly regarded as the RSS at the midpoint time of the time period, thus we need to correct it. To obtain the RSS correction term, it is necessary to compute the difference between the average RSS and RSS at the midpoint over the DFT window. We derive the following proposition for this purpose:

\begin{proposition}[Integral Transform]
	For a continuously differentiable function $f$ on $[a,b]$, the difference between its interval average and its midpoint value can be expressed as:
	\begin{equation}
		\begin{aligned}
			m &=\frac{1}{b-a}\int_{a}^{b}{f}(x)dx-f\left( \frac{a+b}{2} \right)\\
			&=\frac{1}{b-a}\left( \int_{\frac{a+b}{2}}^{b}{(b-t)}{f}'(t)dt-\int_{a}^{\frac{a+b}{2}}{(t-a)}{f}'(t)dt \right)
		\end{aligned}
		\label{integral}
	\end{equation}
	\label{prop2}
\end{proposition}

The proof of Proposition~\ref{prop2} is provided in Section~II of the supplementary material.

\begin{figure}
	\centering
	\includegraphics[width=0.5\textwidth]{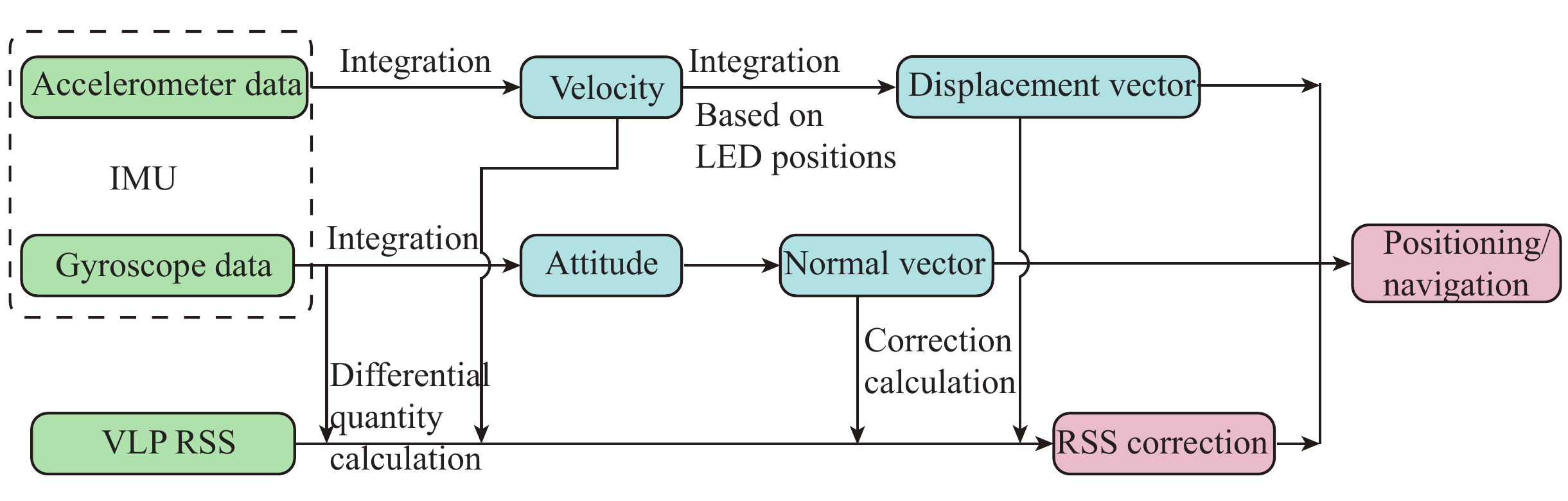}
	\caption{The overall scheme of the RSS correction method.}
	\label{scheme}
\end{figure}

Under the DFT averaging approximation, the RSS correction follows from (\ref{derivative}) and (\ref{integral}):
\begin{align}
& \Delta {{P}_{l}}={{{\tilde{P}}}_{l}}-{{{\bar{P}}}_{l}}={{{\tilde{P}}}_{l}}-\frac{1}{T}\int_{0}^{T}{{{P}_{l}}\mathrm{d}t} \notag\\ 
	& =\frac{1}{T}\int_{0}^{\frac{T}{2}}{-t{{P}_{l}}\frac{\bm{D}_{l}^{u}\times {{\bm{n}}^{u}}}{\bm{D}_{l}^{u}\cdot {{\bm{n}}^{u}}} \cdot\mathrm{d}\boldsymbol{\phi} } \notag\\
	& -\frac{1}{T}\int_{\frac{T}{2}}^{T}{-(T-t){{P}_{l}}\frac{\bm{D}_{l}^{u}\times {{\bm{n}}^{u}}}{\bm{D}_{l}^{u}\cdot {{\bm{n}}^{u}}} \cdot\mathrm{d}\boldsymbol{\phi} } \notag\\ 
	& +\frac{1}{T}\int_{0}^{\frac{T}{2}}{t{{P}_{l}}\left[ -\tfrac{{{\bm{n}}^{u}}}{{{\bm{n}}^{u}}\cdot \bm{D}_{l}^{u}}-\tfrac{{{m}_{l}}\bm{n}_{l}^{u}}{\bm{n}_{l}^{u}\cdot \bm{D}_{l}^{u}}+(3+{{m}_{l}})\tfrac{\bm{D}_{l}^{u}}{D_{l}^{2}} \right] \cdot\mathrm{d}\bm{r}} \notag\\ 
	& -\frac{1}{T}\int_{\frac{T}{2}}^{T}{(T-t){{P}_{l}}\left[ -\tfrac{{{\bm{n}}^{u}}}{{{\bm{n}}^{u}}\cdot \bm{D}_{l}^{u}}-\tfrac{{{m}_{l}}\bm{n}_{l}^{u}}{\bm{n}_{l}^{u}\cdot \bm{D}_{l}^{u}}+(3+{{m}_{l}})\tfrac{\bm{D}_{l}^{u}}{D_{l}^{2}} \right] \cdot\mathrm{d}\bm{r}}
\label{integral_correction}
\end{align}
where ${{\tilde{P}}_{l}}$ represents the instantaneous midpoint RSS over the DFT time window from $l$ LEDs, and ${{\bar{P}}_{l}}$ represents the averaged RSS after integration from $l$ LEDs (i.e., the RSS obtained by DFT). $d\boldsymbol{\phi} $ and $\mathrm{d}\bm{r}$ represent the differentials of attitude angle and position, respectively, both of which change dynamically over time. This expression is in continuous form and cannot be computed by a computer. As shown in Fig. \ref{time_fig}, the discrete Fourier transform window is around the time instant $t_k$, and in discrete form, we use angular rate ${{\boldsymbol{\omega} }}_{i}$ and integrated velocity $\bm{v}_{i}$ to refine these differentials. 
\begin{figure}[H]
	\centering
	\includegraphics[width=3in]{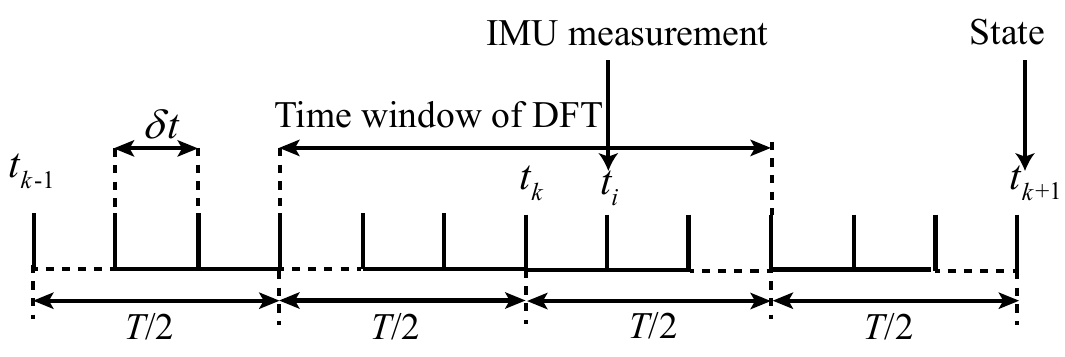}
	\caption{IMU sample time and pre-integration interval.}
	\label{time_fig}
\end{figure}

Therefore, the continuous position differential in (\ref{integral_correction}) is replaced by the displacement increment accumulated over one IMU sampling interval:
\begin{equation}
	\begin{aligned}
	\mathrm{d}\bm{r}&=\bm{v}_{\mathrm{PD}}^u\delta t,\\
	\bm{v}_{\mathrm{PD}}^u&=\bm{v}^u+\bm{C}_v^u\left(\boldsymbol{\omega}^{v}\times\bm{C}_b^v\boldsymbol\ell^{b}\right).
	\end{aligned}
\end{equation}
Supposing the angular rate ${{\boldsymbol{\omega }}}$ obtained from IMU measurements are projected in v-frame, we need to transform it to u-frame:
\begin{equation}
	\mathrm{d}\boldsymbol{\phi}=\bm{C}^u_v\boldsymbol{\omega}^{v}\delta t
\end{equation}
where $\delta t$ is the IMU sampling interval. The integration time changes from 0 to $T$ to ${{t}_{k}}-{T}/{2}$ to ${{t}_{k}}+{T}/{2}$, and taking (\ref{lamber2}) to substitute $P_l$, (\ref{integral_correction}) becomes:
\begin{align}
& \Delta {{P}_{l}}\approx\frac{1}{T}\sum\limits_{t_i=t_k-T/2}^{t_k}{\left( {{t}_{i}}-{{t}_{k}}+\frac{T}{2} \right) \bm{K}_{1,l,i}\cdot\boldsymbol{\omega}_{i}^u\delta t} \notag\\
	& \text{         }-\frac{1}{T}\sum\limits_{t_i=t_k}^{t_k+T/2}{\left( {{t}_{k}}+\frac{T}{2}-{{t}_{i}} \right) \bm{K}_{1,l,i}\cdot\boldsymbol{\omega}_{i}^u\delta t} \notag\\ 
	& \text{         }+\frac{1}{T}\sum\limits_{t_i=t_k-T/2}^{t_k}{\left( {{t}_{i}}-{{t}_{k}}+\frac{T}{2} \right) \bm{K}_{2,l,i}\cdot\bm{v}_{\mathrm{PD},i}^{u}\delta t} \notag\\ 
	& \text{         }-\frac{1}{T}\sum\limits_{t_i=t_k}^{t_k+T/2}{\left( {{t}_{k}}+\frac{T}{2}-{{t}_{i}} \right) \bm{K}_{2,l,i}\cdot\bm{v}_{\mathrm{PD},i}^{u}\delta t}
\label{dP}
\end{align}
where $\bm{K}_1$ and $\bm{K}_2$ represent the partial derivatives of RSS with respect to attitude and position, respectively:
\begin{equation}
	\label{K1}
	\bm{K}_1 = -\frac{\left(m_l+1\right) A_\mathrm{R} P_{\mathrm{T}l}}{2 \mathrm{\uppi}} \cdot \frac{\left(\bm{n}_{l}^u\cdot\bm{D}_l^u\right)^{m_l} \bm{D}_{l}^{u}\times {{\bm{n}}^{u}}} {D^{3+m_l}_l}
\end{equation}
\begin{equation}
	\label{K2}
	\begin{aligned}
	\bm{K}_2 = &\frac{\left(m_l+1\right) A_\mathrm{R} P_{\mathrm{T}l}}{2 \mathrm{\uppi}} \cdot \left[ -\frac{\left(\bm{n}_{l}^u\cdot\bm{D}_l^u\right)^{m_l}{{\bm{n}}^{u}}}{D^{3+m_l}_l} \right.\\
	&-\frac{{{m}_{l}} \left(\bm{n}^u\cdot\bm{D}_l^u\right) \left(\bm{n}_{l}^u\cdot\bm{D}_l^u\right)^{m_l-1} \bm{n}_{l}^{u}} {D^{3+m_l}_l} \\
	&\left.+ \frac{ (3+{{m}_{l}})\left(\bm{n}^u\cdot\bm{D}_l^u\right) \left(\bm{n}_{l}^u\cdot\bm{D}_l^u\right)^{m_l} \bm{D}_{l}^{u}} {D^{5+m_l}_l} \right]
	\end{aligned}
\end{equation}

In (\ref{dP}), $\bm{K}_{1,l,i}$ and $\bm{K}_{2,l,i}$ denote (\ref{K1}) and (\ref{K2}) evaluated for LED $l$ at the IMU timestamp $t_i$, which means $\bm{D}_{l,i}^u$, $\bm{n}_{i}^{u}$, $\bm{K}_{1,l,i}$, and $\bm{K}_{2,l,i}$ are updated at each sample. Variables with subscripts containing time $i$ need to be calculated based on the IMU timestamp (the relationship between times $i$ and $k$ is shown in Fig. \ref{time_fig}). These sample-wise updates account for changes in optical geometry during highly dynamic motion.\footnote{In the reported simulations, using fixed coefficients within the window can introduce approximately 80\% correction error at a translational speed of $2~\mathrm{m/s}$ and approximately 90\% correction error at an angular velocity of $45^\circ/\mathrm{s}$.} In the implemented correction, these pose-, velocity-, and angular-rate-related quantities are initialized from the latest available navigation state and then propagated to the IMU timestamps inside the Fourier window using bias-corrected IMU integration.

The angular velocity and velocity within the Fourier time window are calculated based on the state variables and IMU data. Specifically, the angular velocity can be directly obtained from the gyroscope observations after applying bias correction:

\begin{equation}
\boldsymbol{\omega}_{i}^{u}=\bm{C}_{v_i}^{u}\left(\hat{\boldsymbol{\omega}}_{i}^{v}-\bm{b}_{g}^{v}\right)
\end{equation}
Velocity propagation uses bias-corrected accelerometer specific force and gravity. To first order in the IMU sampling interval,
\begin{equation}
\bm{v}_{i+1}^{u}=\bm{v}_{i}^{u}+\left[\bm{C}_{v_i}^{u}\left(\hat{\bm{a}}_{i}^{v}-\bm{b}_{a}^{v}\right)+\bm{g}^{u}\right]\delta t,
\label{eq:velocity_propagation}
\end{equation}
where $\bm{C}_{v_i}^{u}$ is the attitude-derived DCM from $v$-frame to $u$-frame at $t_i$, $\delta t$ is the IMU sampling interval (Fig.~\ref{time_fig}), and $\hat{\bm{a}}_{i}^{v}$ and $\bm{b}_{a}^{v}$ are the accelerometer specific-force measurement and bias after transformation from $b$-frame by $\bm{C}_{b}^{v}$. The gyroscope measurements and biases above are also expressed in $v$-frame. The gravity vector $\bm{g}^{u}$ points downward in the indoor frame; for an upward-positive $z$-axis, $\bm{g}^{u}=[0,0,-g]^{\mathrm T}$ with $g>0$.

In (\ref{dP})-(\ref{K2}), the position vector \( \bm{D}_{l,i}^u \) and the normal vector ${\bm{n}}_{i}^{u}$ are propagated using IMU measurements for higher temporal resolution within the Fourier window \( T \): the PD trajectory is integrated from velocity, and the attitude is updated by integrating angular rate.

Finally, we use equation (\ref{dP}) to correct the RSS measurements:
\begin{equation}
	P_{l}^{corr}=\bar{P}_{l}+\Delta {{P}_{l}}
\end{equation}
where $\bar{P}_{l}$ is the uncorrected RSS.

An RSS observation assigned to the midpoint $t_k$ of a symmetric DFT window requires samples through $t_k+T/2$. With $T=1.0$ s, this introduces a $0.5$ s midpoint-measurement waiting delay due to data acquisition, separate from computation time. Once the complete window is available, the time-aligned PD and IMU data are used for a single RSS correction before fusion; the corrected observations are held fixed during the subsequent solve.

For each LED and each VLP epoch, the correction in (\ref{dP}) requires four weighted accumulations over the IMU samples inside the DFT window and the evaluation of two local sensitivity vectors, $\bm{K}_1$ and $\bm{K}_2$. If there are $N$ visible LEDs and $M$ IMU samples in the window, the complexity is $O(NM)$ with no additional graph states, residual blocks, or Jacobian terms.

\endgroup

\section{Results and Analysis}
In this section, we evaluate whether correcting the RSS formation stage provides benefits beyond the VLP/INS backend of \cite{sun2025sana}. We report pose errors against the specified references, RSS correction accuracy, sensitivity to translational and rotational motion intensity, DFT window selection, comparison with alternative RSS extraction methods, and physical validation using LiDAR/IMU and independent total-station position references. The data are openly accessible on IEEE DataPort\footnote{Available at: \url{https://dx.doi.org/10.21227/jrwr-rc05}}.

In this study, motion is classified as highly dynamic when the translational speed is at least $1~\mathrm{m/s}$ or the angular speed exceeds $30^\circ/\mathrm{s}$; satisfying either condition is sufficient. This definition applies to both simulations and physical experiments and includes rapid inclination changes at a fixed receiver position. The RSS correction addresses the discrepancy between window-averaged and midpoint RSS, whose magnitude depends on receiver motion, changes in the PD normal, LED geometry, and window length.

\subsection{Simulation}
\label{simulation}

\subsubsection{Simulation setup}
\label{sim:setup}
We built a simulation environment similar to that of \cite{sun2025sana} so that the effect of the new RSS correction layer can be isolated from changes in the fusion backend. For all scenarios, we assume that each LED emits the same optical power with a Lambertian order of 1. In this platform, vehicles are equipped with a PD and an IMU. The Lambertian constant coefficient $\frac{\left(m_l+1\right) A_\mathrm{R} P_{\mathrm{T}l}}{2 \mathrm{\uppi}}$ in (\ref{lamber2}) was set to 500, and the photoelectric signal conversion was not considered in the experiments. To make the data as realistic as possible, we included light reflection effects, PD measurement noise, and ADC quantization noise in the simulated optical measurements. Specifically, 12000 Hz PD current data are generated by a simulated 12-bit ADC that provides integer data ranging from 0 to 4095. Then, we use the DFT algorithm to obtain the RSS and correct it using our proposed method. In the following simulation tests and experiments, the unit of RSS is the least significant bit (LSB) under microcontroller digital signal sampling.

The main-text simulations examine two complementary motion patterns to validate the RSS correction model and its sensitivity to translation and rotation:
\begin{itemize}
    \item \textbf{Scenario 1 (Aerial Dynamics):} Unmanned Aerial Vehicle (UAV) performing circular motion with constant speed while ascending, testing RSS averaging during translational motion. The scenario size is $5\ \mathrm{m} \times 5\ \mathrm{m} \times 5\ \mathrm{m}$. Five LEDs are deployed on the ceiling with planar coordinates of [0, 0], [0, 5], [2.5, 2.5], [5, 5], and [5, 0] (m), respectively.
    
    \item \textbf{Scenario 2 (Handheld Device):} Device at a fixed position with rapid inclination changes, testing RSS averaging caused by changes in the PD normal. The scenario size is the same as scenario 1.
\end{itemize}

\subsubsection{RSS correction model validation}
\label{sim:model_validation}
\begin{figure}[!t]
	\centering
	\includegraphics[width=0.8\columnwidth]{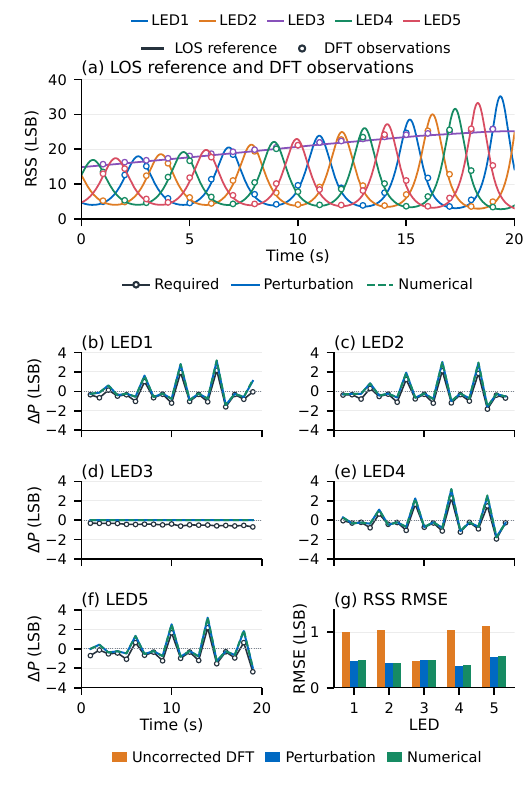}
	\caption{RSS observations and correction in Scenario 1: (a) instantaneous LOS reference (lines) and DFT observations (circles); (b)--(f) required correction and predictions from the perturbation model and numerical Lambertian integration for LEDs 1--5; (g) RSS RMSE for uncorrected DFT, perturbation correction, and numerical correction, with a separate three-entry bar legend. The required correction includes reflection, noise, and quantization effects.}
	\label{rss}
\end{figure}

For the RSS-level model validation in Scenario 1, the UAV starts from [2.5, 4.5, 0] m with an initial velocity of [3, 0, -0.1] m/s and a yaw rate of approximately $-1.50$ rad/s. This rerun introduces PD noise with a per-LED standard deviation of 0.3, and integer ADC quantization. Figure~\ref{rss}(a) overlays the continuous line-of-sight (LOS) RSS reference and DFT-derived observations. The observations are updated at 1 Hz using 1 s windows; only the 19 complete centered windows at 1--19 s are used for the correction comparison and error statistics.

Figure~\ref{rss}(b)--(f) compares the required reference correction, defined as midpoint LOS RSS minus extracted DFT RSS, with the perturbation-model prediction in (\ref{dP}) and numerical Lambertian integration for LEDs 1--5. Both predictions use the same simulated pose and velocity samples at 100 Hz. Numerical integration directly subtracts the window-averaged LOS model from its midpoint value. The two predictions have similar temporal variations, with a mutual RMSE of $0.0246$ LSB. Figure~\ref{rss}(g) compares the three RSS RMSE series explicitly: uncorrected DFT, perturbation correction, and numerical correction. Pooling all five LEDs over the 19 valid epochs gives RMSE values of $0.9671$, $0.4762$, and $0.4870$ LSB, respectively. Differences from the required correction remain because the DFT observations include reflections, noise, and quantization effects. LED3 has a near-zero motion correction and correspondingly little benefit. The Scenario-1 navigation results at the original 2 m/s setting are provided in Section~III of the supplementary material.

Because the correction term in (\ref{dP}) is proportional to the received optical power and the LED transmit power $P_{\mathrm{T}l}$ through the Lambertian model, the absolute correction magnitude decreases under low-power illumination. The relative benefit then depends on the signal-to-noise ratio: if the DFT averaging bias decreases together with RSS but ADC and ambient-light noise remain significant, the corrected RSS can become noise-limited.

\subsubsection{DFT window and IMU sensitivity}
\label{sim:parameter_sensitivity}
\begin{figure}
	\centering
	\includegraphics[width=0.28\textwidth]{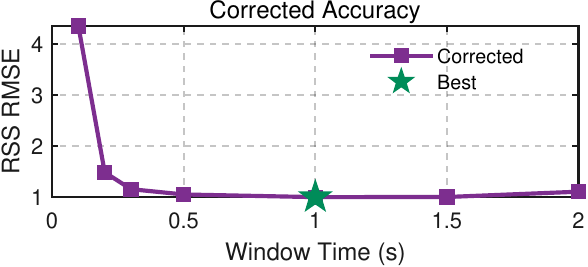}
	\caption{DFT-window-length sensitivity in simulation 1.}
	\label{fig:dft_length}
\end{figure}

Using Scenario 1 at an initial horizontal speed of $2~\mathrm{m/s}$, we extracted and corrected RSS for each DFT window length. Fig.~\ref{fig:dft_length} reports the average RSS error over all visible LEDs. Short windows (0.1 s) give noisy estimates because of limited frequency resolution; at 2.0 s, increased intra-window motion reduces local-linearization accuracy. The error at 1.5 s is only slightly lower than at 1.0 s. We select 1.0 s for its higher RSS update rate and shorter midpoint-measurement waiting delay, balancing spectral separation and motion variation.

To evaluate the sensitivity to IMU quality, we repeated Scenario 1 at an initial horizontal speed of $2~\mathrm{m/s}$ with five grades of simulated IMU errors. The accelerometer velocity random walk $\sigma_{vrw}$, accelerometer bias standard deviation $\sigma_{ba}$, gyroscope angular random walk $\sigma_{arw}$, and gyroscope bias standard deviation $\sigma_{bg}$ are listed in Table~\ref{tab:imu_sensitivity} in SI units. IMU1 represents a high-grade setting, whereas IMU5 represents a noisier MEMS setting. All grades share the same ground-truth trajectory and nominal optical environment. Within each grade, the corrected and uncorrected runs use identical IMU and PD inputs; the optical noise realization is not held identical across grades.

Table~\ref{tab:imu_sensitivity} reports the 3D mean positioning error from the current tightly coupled implementation. RSS correction remains better than the uncorrected baseline for all five IMU grades. The corrected mean error increases from $0.0425$ m for IMU1 to $0.0789$ m for IMU5. The growth at IMU4 and IMU5 indicates sensitivity to noise and bias in the intra-window motion estimates. These results support monitoring IMU uncertainty when the expected correction error becomes comparable to the motion-averaging bias.

\begin{table}[H]
	\caption{IMU noise and bias sensitivity in simulation 1.}
	\label{tab:imu_sensitivity}
	\centering
	\scriptsize
	\setlength{\tabcolsep}{1.5pt}
	\begin{tabular}{ccccccc}
		\hline
		\rule{0pt}{9pt}
		IMU & $\sigma_{vrw}$ & $\sigma_{ba}$ & $\sigma_{arw}$ & $\sigma_{bg}$ & Proposed & \cite{sun2025sana} \\
		 & $(\mathrm{m\,s^{-3/2}})$ & $(\mathrm{m\,s^{-2}})$ & $(\mathrm{rad\,s^{-1/2}})$ & $(\mathrm{rad\,s^{-1}})$ & (m) & (m) \\
		\hline
		IMU1 & $0.0001$ & $0.001$ & $4\times10^{-6}$ & $0.0001$ & $0.0425$ & $0.1430$\\
		IMU2 & $0.0004$ & $0.004$ & $1.6\times10^{-5}$ & $0.0004$ & $0.0443$ & $0.1440$\\
		IMU3 & $0.001$ & $0.01$ & $4\times10^{-5}$ & $0.001$ & $0.0444$ & $0.1433$\\
		IMU4 & $0.004$ & $0.04$ & $1.6\times10^{-4}$ & $0.004$ & $0.0658$ & $0.1502$\\
		IMU5 & $0.01$ & $0.1$ & $4\times10^{-4}$ & $0.01$ & $0.0789$ & $0.1549$\\
		\hline
	\end{tabular}
\end{table}

\subsubsection{Translation and inclination tests}
\label{sim:motion}
\begin{figure}
	\centering
	\begin{minipage}{0.56\linewidth}
		\centering
		\includegraphics[width=\linewidth]{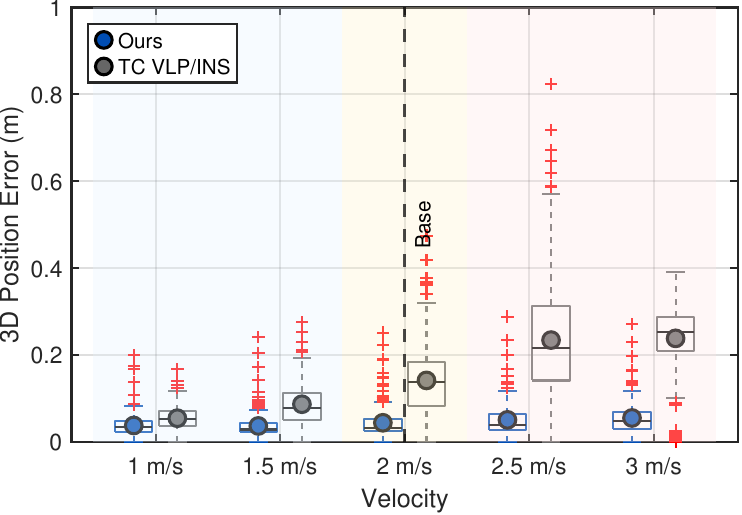}
	\end{minipage}
	\hfill
	\begin{minipage}{0.41\linewidth}
		\centering
		\includegraphics[width=\linewidth]{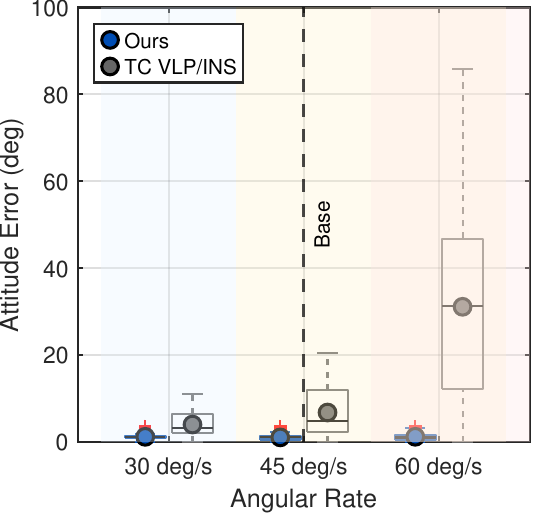}
	\end{minipage}
	\caption{Motion-intensity sensitivity under tightly coupled VLP/INS integration. Circles indicate mean errors. The full uncorrected 60$^\circ$/s output, including its unstable segment, is retained.}
	\label{motion_compare}
\end{figure}

To investigate translational motion intensity, scenario 1 simulated circular ascending motion with initial horizontal speeds of 1, 1.5, 2, 2.5, and 3 m/s and processed each input with the same tightly coupled backend. Fig.~\ref{motion_compare}(a) shows the complete 3D position-error distributions. The mean-error reduction increases from 30.8\% at 1 m/s to 78.4\% at 2.5 m/s, with a slightly smaller reduction of 76.9\% at 3 m/s. The motion inputs were regenerated with random seed 0; their reference trajectories match the original motion settings.

Scenario 2 examined rapid inclination changes at a fixed receiver position. The receiver was initialized at its true starting position $[2.5,4.5,0]$ m. Using identical input data for the two tightly coupled runs, RSS correction reduced the mean absolute pitch error from $4.703^\circ$ to $0.498^\circ$ and the mean absolute roll error from $4.323^\circ$ to $0.875^\circ$.

Additional inclination tests used angular rates of 30, 45, and 60 $^\circ$/s. Fig.~\ref{motion_compare}(b) reports the full distributions of $\sqrt{e_{\rm roll}^2+e_{\rm pitch}^2}$. The corrected mean errors are 1.158, 1.028, and 1.168$^\circ$, compared with 3.985, 6.756, and 31.094$^\circ$ without correction. The uncorrected 60$^\circ$/s run becomes unstable; its complete output is retained in the figure. This failure contributes to the large relative gain at that rate and should not be interpreted as a comparison between two stable navigation solutions.

\subsubsection{RSS extraction methods and offline runtime}
\label{sim:rss_methods}
\begin{figure}[!t]
	\centering
	\includegraphics[width=0.8\columnwidth]{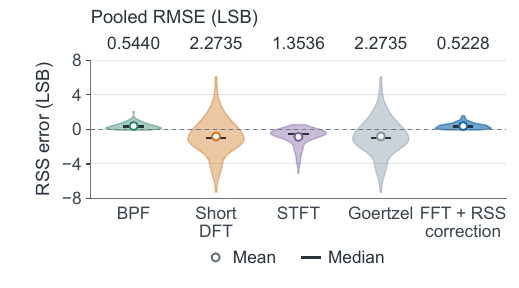}
	\caption{RSS error distributions in simulation 1, pooling five LEDs at 19 valid epochs per LED. Errors are relative to the instantaneous LOS RSS reference. White circles denote means; black segments denote medians. Values above the distributions give pooled RMSE in LSB.}
	\label{fig:rss_methods_error}
\end{figure}

We compared the proposed RSS-corrected FFT with BPF envelope extraction, short-window DFT, STFT, and Goertzel on a separate 20 s Scenario-1 PD signal with an initial horizontal speed of $2~\mathrm{m/s}$, sampled at 12000 Hz.\footnote{The comparison uses random seed 0 and LED tones of 1800, 2500, 3200, 3750, and 4350 Hz. BPF amplitudes are scaled by $1/1.27$ and STFT amplitudes by $1/(1.27\times10.5)$.} BPF used a fourth-order, 2 Hz band-pass filter with zero-phase filtering and Hilbert-envelope extraction. Short-window DFT and Goertzel used 0.50 s blocks (6000 samples); STFT used a 0.0500 s Hann window (600 samples) with 50\% overlap. All methods were evaluated against the instantaneous LOS RSS reference at the same 19 valid epochs, pooling five LEDs to give 95 errors per method. Figure~\ref{fig:rss_methods_error} shows the signed error distributions and pooled RMSE. Corrected FFT achieves the lowest RMSE among the tested configurations, $0.5228$ LSB, compared with $0.5440$ LSB for BPF, $1.3536$ LSB for STFT, and $2.2735$ LSB for both short-window DFT and Goertzel.

For the same 20 s input, median total offline MATLAB R2024b processing times over five repetitions after one warm-up call, with method order rotated, were $0.5637$, $0.0046$, $0.0368$, $0.0034$, and $0.0377$ s for BPF, short-window DFT, STFT, Goertzel, and corrected FFT, respectively. For corrected FFT, $0.0377$ s covers DFT extraction and RSS correction over the full 20 s sequence. These batch processing times do not measure per-update computation time or end-to-end latency. Signal generation, reference preparation, plotting, file output, acquisition waiting, and fusion optimization are excluded.

The simulated positioning-error summary and an additional mixed-motion ground-vehicle scenario are provided in Section~III of the supplementary material. The simulated datasets and RSS correction implementation are available at \url{https://github.com/ShawnSun95/OB_GINS_VLP}.

\subsection{Real-world experiments}
To the best of our knowledge, no existing open VLP dataset covers rapid receiver motion while providing both PD and IMU measurements. We therefore collected our own datasets with synchronized PD and IMU measurements to evaluate the proposed RSS correction and VLP/INS fusion under rapid translation and rotation.

\begin{figure}[!t]
    \centering
    \includegraphics[width=0.75\columnwidth]{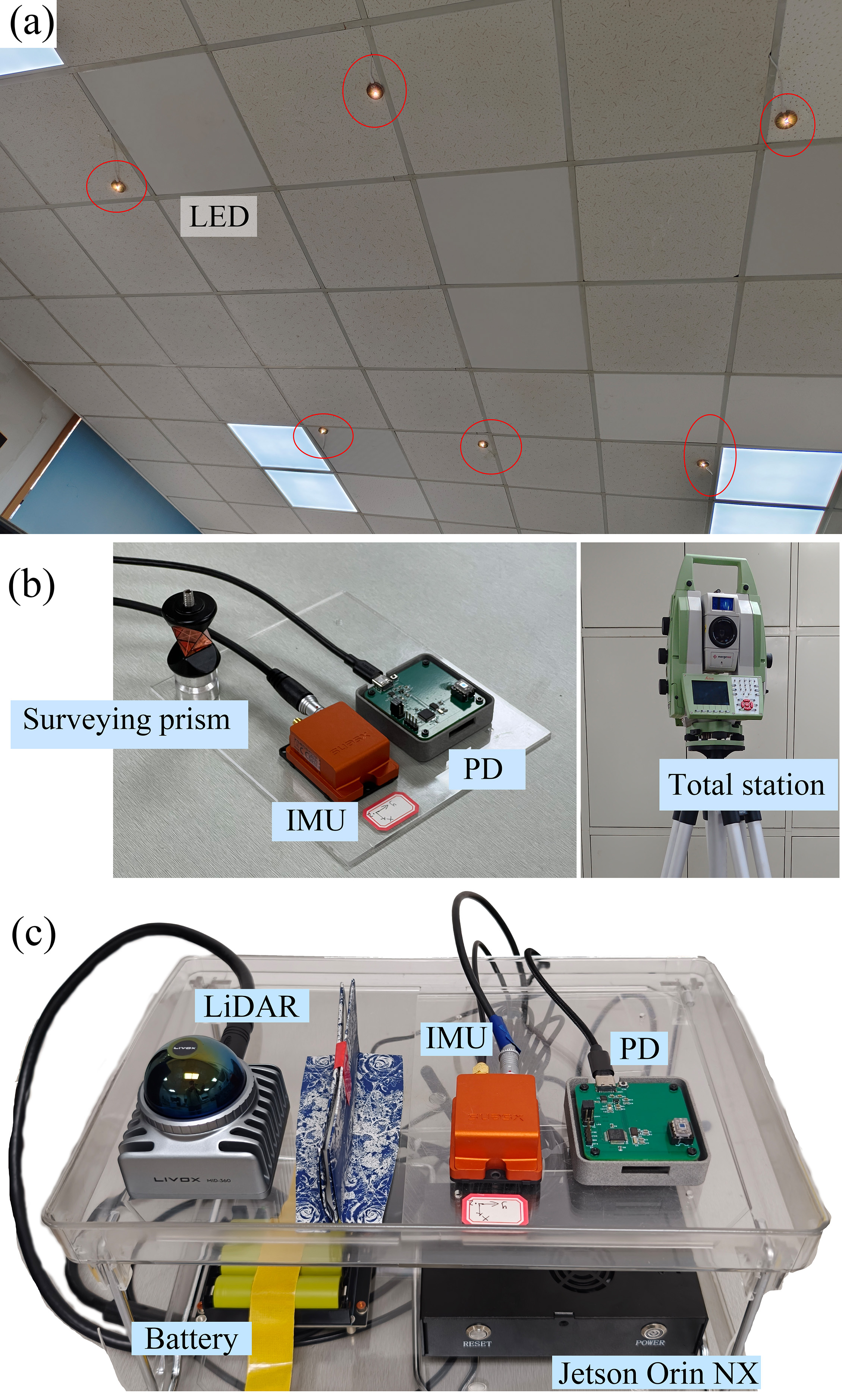}
    \caption{Experimental setups: (a) ceiling-mounted LED layout; (b) handheld terminal and total-station tracking equipment; (c) high-dynamic experimental platform.}
    \label{experiment_scene}
\end{figure}

\begin{figure}[!t]
    \centering
    \includegraphics[width=0.8\columnwidth]{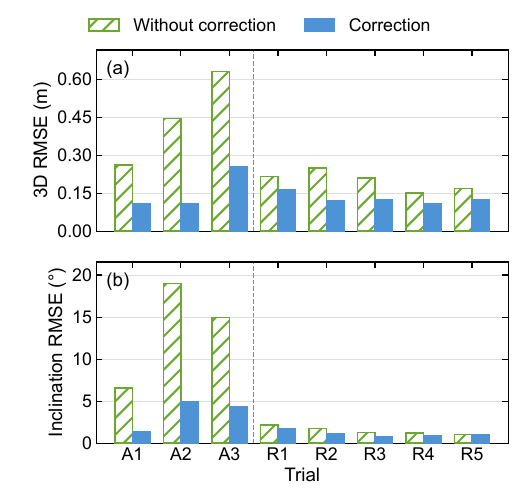}
    \caption{RSS correction in eight held-out trials, evaluated against the LiDAR/IMU reference: (a) 3D position RMSE; (b) PD-normal inclination RMSE. A1--A3: attitude variation; R1--R5: fast running.}
    \label{fig:tc_dynamic_ablation}
\end{figure}

\begin{figure*}[!t]
    \centering
    \includegraphics[width=\textwidth]{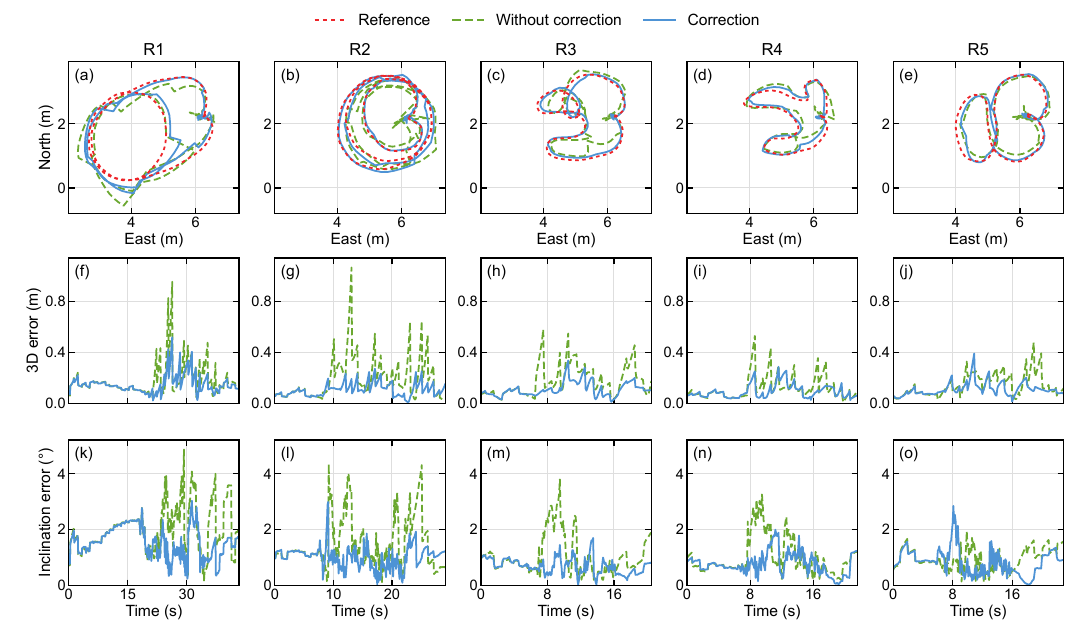}
    \caption{Complete results for all five fast-running trials R1--R5 (columns): (a)--(e) horizontal trajectories; (f)--(j) 3D position errors; (k)--(o) PD-normal inclination errors. All errors are relative to the LiDAR/IMU reference; axes use common scales across trials.}
    \label{fig:tc_dynamic_motion}
\end{figure*}

The physical experiments address two complementary questions: whether RSS correction improves pose estimation under rapid motion, and whether it reduces the discrepancy between DFT-derived RSS and an instantaneous model reference. The total-station trials were collected in November 2025 and the high-dynamic trials in October 2026 in the same indoor laboratory. Substantial changes in the arrangement of desks, chairs, and other furnishings between the two campaigns produced distinct environment configurations. Both campaigns used the same six-LED modulation-frequency configuration (215--865 Hz), an Xsens MTi-G-710 IMU at 200 Hz, and 1 s DFT windows. Total-station measurements provide an independent position reference and, together with IMU attitudes, a Lambertian RSS reference. The LiDAR/IMU reference supports tracking during faster translation and rotation for the high-dynamic pose evaluation. Within each comparison, the correction is switched while the remaining processing settings are fixed. Fig.~\ref{experiment_scene}(a) shows the ceiling-mounted LED layout.

\subsubsection{Pose estimation under rapid motion}
The high-dynamic experimental platform is shown in Fig.~\ref{experiment_scene}(c). An onboard NVIDIA Jetson Orin NX was used for synchronized PD, IMU, and LiDAR data acquisition on the moving experimental platform.

Ten trials were recorded: two static trials, with 10 s dwells, were used exclusively to calibrate LED power and Lambertian order; the eight held-out trials comprised three attitude-variation trials (A1--A3) and five fast-running trials (R1--R5). Their reference-derived 95th-percentile angular speeds and running speeds were $112$--$193^\circ/\mathrm{s}$ and $1.57$--$2.03\ \mathrm{m/s}$, respectively. Peak translational and angular speeds, computed from consecutive reference poses over the same evaluation intervals, reached $2.25~\mathrm{m/s}$ in R2 and $284^\circ/\mathrm{s}$ in A3, respectively. Both variants used the same tightly coupled VLP/INS backend \cite{sun2025sana}, calibration, initialization, and observation weights. The 10 Hz reference used a MID-360 LiDAR, FAST-LIO2 \cite{9697912}, and relocalization against a high-resolution local point-cloud map to suppress accumulated drift. The point-cloud rosbags were replayed at $0.5\times$ speed ($0.25\times$ for R1). Map-frame poses at the IMU origin were retained without per-trial pose resets or additional spatial/time-offset fitting. Because the reference shares the IMU stream with VLP/INS, errors measure reference agreement rather than independent absolute accuracy. Inclination error is the angle between estimated and reference PD normals.

Relative to the LiDAR/IMU reference, correction lowers the 3D RMSE in every held-out trial (Fig.~\ref{fig:tc_dynamic_ablation}). Pooling all 2487 common epochs, the 3D RMSE decreases from $0.329$ to $0.145$ m (56.0\% reduction), the inclination RMSE from $9.24^\circ$ to $2.60^\circ$ (71.9\% reduction), and the 95th-percentile 3D error from $0.634$ to $0.255$ m. Fig.~\ref{fig:tc_dynamic_motion} presents all five running trials. Correction reduces trajectory deviations and large position-error excursions; the pooled 3D RMSE falls from $0.207$ to $0.135$ m (34.9\% reduction), and the 95th-percentile error from $0.395$ to $0.235$ m. For A1--A3, correction reduces pooled roll RMSE from $3.76^\circ$ to $1.25^\circ$ and pitch RMSE from $3.29^\circ$ to $1.39^\circ$; the complete roll and pitch error curves are provided in Section~IV-A of the supplementary material.

\subsubsection{Total-station validation of RSS and positioning}
\begin{figure}[!t]
    \centering
    \includegraphics[width=0.85\columnwidth]{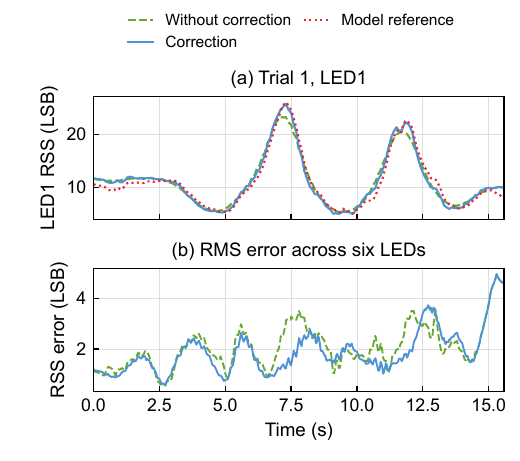}
    \caption{RSS validation in representative Trial 1: (a) LED1 observations and model reference; (b) RMS RSS error over all six LEDs.}
    \label{experiment_RSS}
\end{figure}

Four handheld trials used the setup in Fig.~\ref{experiment_scene}(b), where a Leica MS50 total station tracked a prism rigidly mounted with the PD and Xsens IMU to provide an independent position reference. After time-offset and lever-arm compensation, total-station positions and IMU orientations were used to predict instantaneous RSS with the Lambertian model. RSS RMSE decreases in all four trials. Fig.~\ref{experiment_RSS} illustrates Trial 1, selected by the lower median uncorrected position RMSE: LED1 shows reduced discrepancies near local extrema, while the six-LED error summarizes the full RSS evaluation interval. In these trials, the mean 95th-percentile translation and angular speeds were $1.00\ \mathrm{m/s}$ and $49.6^\circ/\mathrm{s}$.

With the simple integration method in Section~\ref{simple integration}, pooling all 350 common total-station epochs gives a 3D position RMSE decrease from $0.137$ to $0.114$ m (16.8\% reduction) with RSS correction. The evaluation protocol, a representative trajectory, the pooled positioning-error CDF, and per-trial accuracy statistics for these simple-integration trials are provided in Section~IV-B of the supplementary material. The datasets use different motion regimes and fusion backends, so their percentage gains do not directly compare positioning accuracy. Code and data are available in the VLP toolbox\footnote{\url{https://github.com/ShawnSun95/VLPtoolbox}} and the aforementioned IEEE DataPort link.

\section{Conclusion}
We proposed an IMU-aided RSS correction that compensates for the difference between window-averaged DFT measurements and instantaneous RSS during receiver motion. The correction operates before fusion and can be used with both simple and tightly coupled VLP/INS integration. Controlled simulations validate the DFT averaging relation and the correction formula against known RSS references and numerical Lambertian integration. Across eight held-out dynamic trials, the 3D position and PD-normal inclination RMSE relative to the LiDAR/IMU reference decrease by 56.0\% and 71.9\%, respectively. This reference shares the IMU stream with VLP/INS. Independent total-station position measurements confirm a 16.8\% reduction in pooled 3D position RMSE, alongside reduced RSS discrepancies against the Lambertian reference constructed from measured positions and IMU attitudes. These results support correcting motion-induced measurement bias when applying PD-based VLP to rapidly moving indoor platforms.

\FloatBarrier
\bibliographystyle{IEEEtran}
\bibliography{reference}

\raggedend

\vfill
\end{document}